\documentstyle[aps,preprint]{revtex}

\begin{document}
\title{Fermionic zero-norm states and enlarged supersymmetries of Type II string}
\author{Jen-Chi Lee\thanks{%
corresponding author: e-mail: jcclee@cc.nctu.edu.tw; }}
\address{Department of Electrophysics, National Chiao Tung University,\\
Hsinchu 30050, Taiwan, R. O. C}
\date{\today}
\maketitle

\begin{abstract}
We calculate the NS-R fermionic zero-norm states of type II string spectrum.
The massless and some possible massive zero-norm states are identified to be
responsible for the space-time supersymmetry. The existence of other
fermionic massive zero-norm states with higher spinor-tensor indices
correspond to new enlarged boson-fermion symmetries of the theory at high
energy. We also discuss the R-R charges and R-R zero-norm states and justify
that perturbative string does not carry the massless R-R charges. However,
the existence of some massive R-R zero-norm states make us speculate that
string may carry \ some {\it massive} R-R charges.
\end{abstract}

\newpage

\section{\protect\bigskip INTRODUCTION}

It has been known for a while that the complete space-time symmetry of
string theory is closely related to the existence of zero-norm states\cite{1}
in the old covariant quantization of the string spectrum.\cite{2} This
include the $w_{\infty }$ symmetry of 2d Liouville theory and the {\it %
discrete} T-duality symmetry of closed bosonic string.\cite{3} However, all
previous study was about the bosonic sector and nothing has been said about
the symmetries of NS-R( Nevenu Schwars-Ramond) and R-R sectors, ex. NS-R
supersymmetry zero-norm states. On the other hand, recent study has revealed
that D-brane is the symmetry charge carrier of massless R-R state.\cite{4} A
further study from zero-norm state point of view may hold the key to uncover
the whole set of R-R charges including the massive ones.

In this paper, we will identify the fermionic zero-norm states which are
responsible for the space-time supersymmetry in the NS-R sector of Type II
string. This includes the massless and some possible massive ones since
supersymmetry is an exact symmetry at each fixed mass level of the spectrum.
For simplicity, we will calculate the zero-norm states up to the first
massive level. In addition to the supersymmetry zero-norm states, we
discover other massive zero-norm states with higher spinor-tensor indices
which, presumably, correspond to new enlarged massive boson-fermion
symmetries of the theory. We then calculate the massless R-R zero-norm
states. It is found that they have exactly the same degree of freedom with
those of the positive-norm physical propagating R-R gauge fields, and thus
do not fit into the R-R charge degree of freedom. We conclude that the
massless R-R zero-norm states {\it do not }correspond to the charges of
massless R-R gauge fields. There is no perturbative string states which
carries the massless R-R charges. This is very different from the NS-NS and
NS-R sectors. A further study of the meaning of these massless R-R zero-norm
states seems necessary. On the other hand, the existence of massive R-R
zero-norm states may imply that some massive perturbative string R-R states
do carry some kinds of {\it massive} R-R charges. We propose a method to
justify this important issue.

This paper is organized as following. In section II, we calculate the NS
zero-norm states including the first massive even G-parity states. In
section III, we calculate the massive R-states and discover a type I
massless fermionic zero-norm state and two types of massive zero-norm
R-states. Section IV is devoted to the discussion of physical meaning of all
kinds of NS-R and R-R zero-norm states. In particular, supersymmetry
zero-norm states are identified. Finally, we give a brief discussion and
make some speculations about the massive R-R charges.

\section{NEVEU-SCHWARZ STATES}

In this section we work out all the physical states (including two types of
bosonic zero-norm states) of the spectrum after GSO projection\cite{5} in
the NS sector of the open superstring. For the massless states (We use the
notation in Ref.\cite{1}),

\begin{equation}
\varepsilon _{\mu }b_{-1/2}^{\mu }\left| 0,k\right\rangle ;k\cdot
\varepsilon =0.  \eqnum{2.1}
\end{equation}
The first massive states ($-k^{2}=m^{2}=2$) are

\begin{equation}
\varepsilon _{\mu \nu \lambda }b_{-1/2}^{\mu }b_{-1/2}^{\nu
}b_{-1/2}^{\lambda }\left| 0,k\right\rangle ;\varepsilon _{\mu \nu \lambda
}\equiv \varepsilon _{\left[ \mu \nu \lambda \right] },k_{{}}^{\mu
}\varepsilon _{\mu \nu \lambda }=0  \eqnum{2.2}
\end{equation}
and

\begin{equation}
\varepsilon _{\mu \nu }\alpha _{-1}^{\mu }\alpha _{-1/2}^{\nu }\left|
0,k\right\rangle ;\varepsilon _{\mu \nu }=\varepsilon _{\nu \mu
},k_{{}}^{\mu }\varepsilon _{\mu \nu }=\eta ^{\mu \nu }\varepsilon _{\mu \nu
}=0.  \eqnum{2.3}
\end{equation}

In addition to the above positive-norm states, there are two types of
zero-norm states in the NS sector. They are

Type I:

\begin{equation}
G_{-1/2}\left| \chi \right\rangle ,\text{ where }G_{1/2}\left| \chi
\right\rangle =G_{3/2}\left| \chi \right\rangle =L_{0}\left| \chi
\right\rangle =0,  \eqnum{2.4}
\end{equation}
and

Type II:

\begin{equation}
(G_{3/2}+2G_{-1/2}L_{-1})\left| \chi \right\rangle ,\text{ where }%
G_{1/2}\left| \chi \right\rangle =G_{3/2}\left| \chi \right\rangle
=(L_{0}+1)\left| \chi \right\rangle =0.  \eqnum{2.5}
\end{equation}

Note that type I states have zero-norm at any space-time dimension while
type II states have zero-norm only at D=10. For the massless level, we have
only one singlet type I zero-norm state

\begin{equation}
\left| \chi \right\rangle =\left| 0,k\right\rangle ,-k^{2}=m^{2}=0, 
\eqnum{2.6}
\end{equation}

\begin{equation}
G_{-1/2}\left| \chi \right\rangle =k\cdot b_{-1/2}\left| 0,k\right\rangle . 
\eqnum{2.7}
\end{equation}
For the first massive level (even G-parity, $-k^{2}=m^{2}=2$)

Type I:

1.

\begin{equation}
\left| \chi \right\rangle =\theta _{\mu \nu }b_{-1/2}^{\mu }b_{-1/2}^{\nu
}\left| 0,k\right\rangle ;\theta _{\mu \nu }=-\theta _{\nu \mu },k^{\mu
}\theta _{\mu \nu }=0,  \eqnum{2.8}
\end{equation}

\begin{equation}
G_{-1/2}\left| \chi \right\rangle =[2\theta _{\mu \nu }\alpha _{-1}^{\mu
}b_{-1/2}^{\nu }+k_{[\lambda }\theta _{\mu \nu ]}b_{-1/2}^{\lambda
}b_{-1/2}^{\mu }b_{-1/2}^{\nu }]\left| 0,k\right\rangle .  \eqnum{2.9}
\end{equation}

2.

\begin{equation}
\left| x\right\rangle =\theta _{\mu }\alpha _{-1}^{\mu }\left|
0,k\right\rangle ,k\cdot \theta =0,  \eqnum{2.10}
\end{equation}

\begin{equation}
G_{-1/2}\left| \chi \right\rangle =[\theta \cdot b_{-3/2}+(k\cdot
b_{-1/2})(\theta \cdot \alpha _{-1})]\left| 0,k\right\rangle .  \eqnum{2.11}
\end{equation}

3.

\begin{equation}
\left| \chi \right\rangle =[\theta \cdot \alpha _{-1}+(k\cdot
b_{-1/2})(\theta \cdot b_{-1/2})]\left| 0,k\right\rangle ,k\cdot \theta =0, 
\eqnum{2.12}
\end{equation}

\begin{equation}
G_{-1/2}\left| \chi \right\rangle =\{2(k\cdot b_{-1/2})(\theta \cdot \alpha
_{-1})+(k\cdot \alpha _{-1})(\theta \cdot b_{-1/2})+\theta \cdot
b_{-2/3}\}\left| 0,k\right\rangle .  \eqnum{2.13}
\end{equation}
The discovery of state (2.13) is suggested by noting that state (2.12) with $%
-k^{2}=m^{2}=1$ correspond to odd G-parity zero-norm state.

Type II:

\begin{equation}
\left| \chi \right\rangle =\left| 0,k\right\rangle ,  \eqnum{2.14}
\end{equation}

\begin{equation}
(G_{-3/2}+2G_{-1/2}L_{-1})\left| \chi \right\rangle =\{3k\cdot
b_{-2/3}+2(k\cdot b_{-1/2})(k\cdot \alpha _{-1})+(b_{-1/2}\cdot \alpha
_{-1})\}\left| 0,k\right\rangle .  \eqnum{2.15}
\end{equation}
The NS-NS sector of Type II string spectrum can be obtained by

\begin{equation}
\left| \phi \right\rangle =\left| \phi \right\rangle _{R}\otimes \left| \phi
\right\rangle _{L}  \eqnum{2.16}
\end{equation}
where $\left| \phi \right\rangle $ is zero-norm if either $\left| \phi
\right\rangle _{R}$ or $\left| \phi \right\rangle _{L}$ is zero-norm. As in
the case of bosonic string, each zero-norm state corresponds to a symmetry
transformation on the bosonic sector.

\section{RAMOND STATES}

We now discuss the interesting R-sector. The zero-mode of worldsheet
fermionic operators $d_{0}^{\mu }=\frac{-i}{\sqrt{2}}\Gamma ^{\mu },$ where $%
\Gamma ^{\mu }$ obeys the 10D Dirac algebra

\begin{equation}
\{\Gamma ^{\mu },\Gamma ^{\nu }\}=-2\eta ^{\mu \nu }.  \eqnum{3.1}
\end{equation}

Define \cite{6}

\begin{eqnarray}
\Gamma ^{a\pm } &=&\frac{1}{\sqrt{2}}(d_{0}^{2a}\pm
id_{0}^{2a+1}),(a=1,2,3,4)  \nonumber \\
\Gamma ^{0\pm } &=&\frac{1}{\sqrt{2}}(d_{0}^{1}\mp d_{0}^{0}),  \eqnum{3.2}
\end{eqnarray}
then

\begin{equation}
\{\Gamma ^{a+},\Gamma ^{b-}\}=\delta ^{ab},(a,b=0,1,2,3,4).  \eqnum{3.3}
\end{equation}
That is $\Gamma ^{a\pm }$ are raising and lowering operators. The ground
state of the R-sector can be labeled by

\begin{equation}
\left| \overrightarrow{S}\right\rangle \equiv \left|
S_{0},S_{1},S_{2},S_{3},S_{4}\right\rangle \equiv \left| \overrightarrow{S}%
,k\right\rangle u_{\overrightarrow{s}}  \eqnum{3.4}
\end{equation}
with $S_{a}=\pm \frac{1}{2},$ $a=0,1,2,3,4$ ,and $u_{\overrightarrow{s}}$ in
equation (3.4) is the spin polarization. The 32 off-shell states decompose $%
32\rightarrow 16+16^{\prime }$ according to even (odd) numbers of -1/2 on $%
S_{a}.$ The on-shell physical state conditions are

\begin{equation}
F_{0}\left| \overrightarrow{S},k\right\rangle u_{\overrightarrow{s}}=0 
\eqnum{3.5}
\end{equation}
and

\begin{equation}
F_{1}\left| \overrightarrow{S},k\right\rangle u_{\overrightarrow{s}}=0. 
\eqnum{3.6}
\end{equation}

Equation (3.6) is automatically satisfied for the state in equation (3.4)
and equation (3.5) implies the massless Dirac equation $k_{\mu }\Gamma _{%
\overrightarrow{s}^{\prime }\overrightarrow{s}}^{\mu }u_{\overrightarrow{s}%
}=0.$ In the frame $k^{0}=k^{1},k^{i}=0,i=2,3,...9,$ this implies $S_{0}=%
\frac{1}{2}.$ The 16 on-shell states decompose again $16\rightarrow
8_{s}+8_{c}\equiv u_{\overrightarrow{s}}+\overline{u}_{\overrightarrow{s}}.$
At the massless level, the GSO operator reduces to the chirality operator,
and only one of the chiral spinor $8_{s}$(or $8_{c}$) will be projected out.
There are two massive vector-spinor states:

1.

\begin{equation}
\alpha _{-1}^{\mu }\left| \overrightarrow{S},k\right\rangle u_{\mu ,%
\overrightarrow{s}}.  \eqnum{3.7}
\end{equation}
The physical state conditions (3.5) and (3.6) give

\begin{equation}
\lbrack (k\cdot d_{0})\alpha _{-1}^{\mu }+d_{-1}^{\mu }]\left| 
\overrightarrow{S},k\right\rangle u_{\mu ,\overrightarrow{s}}=0,  \eqnum{3.8}
\end{equation}
and

\begin{equation}
d_{0}^{\mu }\left| \overrightarrow{S},k\right\rangle u_{\mu ,\overrightarrow{%
s}}=0.  \eqnum{3.9}
\end{equation}

2.

\begin{equation}
d_{-1}^{\mu }\left| \overrightarrow{S},k\right\rangle \overline{u}_{\mu ,%
\overrightarrow{s}}.  \eqnum{3.10}
\end{equation}
The physical state conditions (3.5) and (3.6) give

\begin{equation}
\lbrack (k\cdot d_{0})d_{-1}^{\mu }+\alpha _{-1}^{\mu }]\left| 
\overrightarrow{S},k\right\rangle \overline{u}_{\mu ,\overrightarrow{s}}=0, 
\eqnum{3.11}
\end{equation}
and

\begin{equation}
k^{\mu }\left| \overrightarrow{S},k\right\rangle \overline{u}_{\mu ,%
\overrightarrow{s}}=0.  \eqnum{3.12}
\end{equation}
Note that equation (3.8) and (3.11) imply $-k^{2}=m^{2}=2$ since $%
L_{0}=F_{0}^{2}.$

In addition to the above positive-norm states, there are two types of
zero-norm states in the R-sector. They are

Type I:

\begin{equation}
F_{0}\left| \psi \right\rangle ,\text{ where }F_{1}\left| \psi \right\rangle
=L_{0}\left| \psi \right\rangle =0;  \eqnum{3.13}
\end{equation}
and

Type II:

\begin{equation}
F_{0}F_{-1}\left| \psi \right\rangle ,\text{ where }F_{1}\left| \psi
\right\rangle =(L_{0}+1)\left| \psi \right\rangle =0.  \eqnum{3.14}
\end{equation}
We have used the superconformal algebra in the R-sector to simplify the
equations in (3.13) and (3.14). Note that, as in the NS-sector, type I
states have zero-norm at any space-time dimension while type II states have
zero-norm only at D=10. For the massive level, we have only one type I
zero-norm state

\begin{equation}
\left| \psi \right\rangle =\left| \overrightarrow{S},k\right\rangle \theta _{%
\overrightarrow{s}},-k^{2}=m^{2}=0,  \eqnum{3.15}
\end{equation}

\begin{equation}
F_{0}\left| \overrightarrow{S},k\right\rangle \theta _{\overrightarrow{s}}=%
\frac{1}{i\sqrt{2}}k_{\mu }\Gamma _{\overrightarrow{s}^{\prime }%
\overrightarrow{s}}^{\mu }\left| \overrightarrow{S},k\right\rangle \theta _{%
\overrightarrow{s}}.  \eqnum{3.16}
\end{equation}
Note that $(k\cdot \Gamma )(k\cdot \Gamma )\left| \overrightarrow{S}%
,k\right\rangle \theta _{\overrightarrow{s}}=0$ implies, in the $k^{0}=k^{1}$
frame, $S_{0}^{\prime }=1/2.$ The spinor in equation (3.16) has 16 on-shell
states as the spinor in equation (3.4). In particular, the spinor $(k\cdot
\Gamma )\left| \overrightarrow{S},k\right\rangle \theta _{\overrightarrow{s}%
} $ is left-handed if $\left| \overrightarrow{S},k\right\rangle \theta _{%
\overrightarrow{s}}$ is right-handed and vice versa. For the type I massive
level:

1.

\begin{equation}
\left| \psi \right\rangle =\alpha _{-1}^{\mu }\left| \overrightarrow{S}%
,k\right\rangle \theta _{\mu ,\overrightarrow{s}},-k^{2}=m^{2}=2, 
\eqnum{3.17}
\end{equation}

\begin{equation}
F_{1}\left| \psi \right\rangle =d_{0}^{\mu }\left| \overrightarrow{S}%
,k\right\rangle \theta _{\mu ,\overrightarrow{s}}=0,  \eqnum{3.18}
\end{equation}

\begin{equation}
F_{0}\left| \psi \right\rangle =[(k\cdot d_{0})\alpha _{-1}^{\mu
}+d_{-1}^{\mu }]\left| \overrightarrow{S},k\right\rangle \theta _{\mu ,%
\overrightarrow{s}}.  \eqnum{3.19}
\end{equation}

2.

\begin{equation}
\left| \psi \right\rangle =d_{-1}^{\mu }\left| \overrightarrow{S}%
,k\right\rangle \overline{\theta }_{\mu ,\overrightarrow{s}},-k^{2}=m^{2}=2,
\eqnum{3.20}
\end{equation}

\begin{equation}
F_{1}\left| \psi \right\rangle =k_{{}}^{\mu }\left| \overrightarrow{S}%
,k\right\rangle \overline{\theta }_{\mu ,\overrightarrow{s}}=0,  \eqnum{3.21}
\end{equation}

\begin{equation}
F_{0}\left| \psi \right\rangle =[(k\cdot d_{0})d_{-1}^{\mu }+\alpha
_{-1}^{\mu }]\left| \overrightarrow{S},k\right\rangle \overline{\theta }%
_{\mu ,\overrightarrow{s}}.  \eqnum{3.22}
\end{equation}
Equation (3.18) and (3.21) are gauge conditions. Note that (3.7), (3.10),
(3.19) and (3.22) can be decomposed into $56_{s}+8_{c}$ or $56_{c}+8_{s}$ in
the light-cone gauge. There is one type II zero-norm state at this massive
level

\begin{equation}
\left| \psi \right\rangle =\left| \overrightarrow{S},k\right\rangle \theta _{%
\overrightarrow{s}},-k^{2}=m^{2}=2,  \eqnum{3.23}
\end{equation}

\begin{equation}
F_{0}F_{-1}\left| \psi \right\rangle =[d_{-1}\cdot d_{0}+k\cdot \alpha
_{-1}+(k\cdot d_{0})(k\cdot d_{-1})+(k\cdot d_{0})(\alpha _{-1}\cdot
d_{0})]\left| \overrightarrow{S},k\right\rangle \theta _{\overrightarrow{s}}.
\eqnum{3.24}
\end{equation}

\section{SYMMETRY CHARGES AND\ ZERO-NORM STATES\ OF\ NS-R AND\ R-R SECTORS}

\subsection{ NS-R sector}

The fermionic zero-norm states can be obtained by equation (2.16), in which
either $\left| \psi \right\rangle _{R}(\left| \psi \right\rangle _{L})$ is a
NS-zero-norm state or $\left| \psi \right\rangle _{R}(\left| \psi
\right\rangle _{L})$ is a R-zero-norm state. We first identify the massless
supersymmetry zero-norm states which are responsible for the N=2 massless
space-time supersymmetry transformation. The obvious candidates are the
products of states in equations (2.7) and (3.4)

\begin{equation}
k\cdot b_{-1/2}\left| 0,k\right\rangle \otimes \left| \overrightarrow{S}%
,k\right\rangle u_{\overrightarrow{s}},  \eqnum{4.1}
\end{equation}
and

\begin{equation}
\left| \overrightarrow{S},k\right\rangle \overline{u}_{\overrightarrow{s}%
}\otimes k\cdot \overline{b}_{-1/2}\left| 0,k\right\rangle  \eqnum{4.2}
\end{equation}
for the type II A string. For II B string, the spinors in equations (4.1)
and (4.2) are chosen to have the same chirality. Other possible choices is
to use the zero-norm state in equation (3.16) on the R-sector side, and
either state in equation (2.1) or (2.7) on the NS-sector side. For the case
of equation (2.1), one notes that we have the tensor-spinor decomposition

\begin{equation}
8_{v}\otimes 8_{c}=56_{c}\oplus 8_{s},  \eqnum{4.3}
\end{equation}

\begin{equation}
8_{v}\otimes 8_{s}=56_{s}\oplus 8_{c}  \eqnum{4.4}
\end{equation}
where $56_{s,c}$ are gravitons. In this case, though we get the right
supersymmetry spinor charge indices $8_{c,s},$ the $56_{s,c}$ zero-norm
states do not correspond to any symmetry. One also notes that, as stated in
the paragraph below equation (3.16), the states $k\cdot \Gamma \left| 
\overrightarrow{S},k\right\rangle \theta _{\overrightarrow{s}}$ has exactly
the same degree of freedom with the positive-norm state $\left| 
\overrightarrow{S},k\right\rangle u_{\overrightarrow{s}}$ and also $u_{%
\overrightarrow{s}},$ $\overline{\theta }_{\overrightarrow{s}}$ have
different chiralities. We conclude that the massless R-zero-norm state
(3.16) {\it does not} correspond to the space-time symmetry. Similar
consideration will be discussed in the R-R sector.

We now discuss the massive zero-norm states. We first identify zero-norm
states with only one space-time spinor index. They are all possible
candidates for the massive space-time supersymmetry zero-norm states. For
example, one can choose

\begin{equation}
(2.15)\otimes 8_{c,s}  \eqnum{4.5}
\end{equation}
where (2.15) is the singlet type II zero-norm state in equation (2.15), and $%
8_{c,s}$ are the components of decomposition through equations (4.3) or
(4.4) of any one of equations (3.7), (3.10), (3.19) and (3.22). Instead of
using states (2.15) in equation (4.5), one can choose either the vector
state in equation (2.11) or equation (2.13), and then make use of the
decomposition (4.3) or (4.4) again to get the spinor $8_{c,s}$ states. We
are unable to identify which states stated above do correspond to the
massive supersymmetry transformation. However, the massive supersymmetry
zero-norm states are surely among the set we stated above.

In addition to the massive spinor $8_{c,s}$ zero-norm states, we have many
other fermionic massive zero-norm states with higher spinor-tensor indices
which, presumably, will generate many new enlarged stringy boson-fermion
supersymmetries. This is an analogy of the enlarged bosonic gauge symmetries
in the NS-NS sector discussed in Ref.\cite{2}. We give one example here

\begin{equation}
\theta _{\lbrack \mu \nu ]}\otimes 56_{c,s},  \eqnum{4.6}
\end{equation}
where $\theta _{\lbrack \mu \nu ]}$ is the zero-norm state in equation (2.9)
and $56_{c,s}$ is the positive-norm state in either (3.7) or (3.10).
Equation (4.6) represents the charge of a complicated inter-spin symmetry of
stringy physics.

\subsection{ R-R sector}

The massless R-R states of type II string consist of antisymmetric tensor
forms

\begin{equation}
G_{\alpha ,\beta }=\sum_{k=0}^{10}\frac{i^{k}}{k!}G_{\mu _{1}\mu _{2}...\mu
_{k}}(\Gamma ^{_{\mu _{1}\mu _{2}...\mu _{k}}})_{\alpha \beta },  \eqnum{4.7}
\end{equation}
where $\Gamma ^{_{\mu _{1}\mu _{2}...\mu _{k}}}$ are the antisymmetric
products of gamma-matrix, and $\alpha ,\beta $ are spinor indices. There is
a duality relation which reduces the number of independent tensor components
to up to k=5 form. The on-shell conditions, or two massless Dirac equations,
imply

\begin{equation}
k^{[\mu }G^{_{\nu _{1}...\nu _{k}}]}=k_{\mu }G^{\mu \nu _{2}...\nu _{k}}=0, 
\eqnum{4.8}
\end{equation}
which means $G$ is a {\it field strength }and can be written as a total
derivative

\begin{equation}
G_{(k)}=dA_{(k-1)}.  \eqnum{4.9}
\end{equation}

Equation (4.9) has one important consequence that perturbative string states
do not carry the {\it massless} R-R symmetry charges. It was argued that the
nonperturbative black p-branes\cite{7} carry the R-R charges before
Polchinski realized that D-branes\cite{4} are exact string soliton states
which carry the massless R-R charges. It this subsection, we will discuss
this issue from zero-norm state point of view. For the massless level, we
have the following zero-norm states

\begin{equation}
k_{\mu }\Gamma _{\overrightarrow{s}^{\prime }\overrightarrow{s}}^{\mu
}\left| \overrightarrow{S},k\right\rangle \theta _{\overrightarrow{s}%
}\otimes \left| \overrightarrow{S},k\right\rangle u_{\overrightarrow{s}} 
\eqnum{4.10}
\end{equation}
which decompose into R-R forms according to (4.7). Note that $k\cdot \Gamma
\theta \otimes k\cdot \Gamma \theta =0.$ However, since the number of degree
of freedom of states in equation (4.10) do not fit into that of R-R charges,
and for reasons from the NS-R sector in the paragraph below equation (4.4),
we conclude that zero-norm states in equation (4.10) do not correspond to
symmetry charges. This justifies that perturbative string does not carry the
massless R-R charge.

For the massive level, we have R-zero-norm states in equation (3.19), (3.22)
and (3.23). Note that the open string vector-spinor states (3.19) and (3.22)
can be decomposed into $56_{s,c}\oplus 8_{c,s}$ according to equations (4.3)
and (4.4). By the same argument, relating to the counting of the number of
degree of freedom, we conclude that the $56_{s,c}$ zero-norm states
decomposed from the R-zero-norm states (3.19) and (3.22) do not correspond
to symmetry charges. We also conjecture that {\it the open string massive }$%
8_{s,c}${\it \ zero-norm states decomposed from equations (3.19) and (3.22)
serve as symmetry charges of open string massive }$56_{s,c}${\it \
positive-norm states decomposed from (3.7) and (3.10), and that the open
string massive }$8_{s,c}${\it \ positive-norm states find no symmetry
charges in the perturbative string spectrum as in the case of massless level.%
} Similar statement can be made for the R-R states of closed type II string.

The charges of massive $8_{s,c}$ states may come from D-branes as well. It
seems that for the vector-spinor states, the charges exist only in the
''vector'' part of the states from zero-norm state point of view. According
to our argument, {\it perturbative string does carry some massive R-R charges%
}. In addition to the right number of degree of freedom of zero-norm state
charges, one evidence to support this statement is that we cannot have
D-branes with D$\geq $10 to couple k-tensors with k$\geq $11 which do exist
in the massive R-R spectrum. One way to justify the conjecture stated above
is to repeat the calculation of equations (4.7) to (4.9) for the massive
modes. Work in this direction is in progress.

\section{ACKNOWLEDGEMENTS}

\bigskip I would like to thank Darwin Chang and Pei-Ming Ho for discussions.
This research is supported by National Science Council, Taiwan under the
grant number NSC 88-2112-M-009-011.

\end{document}